# Solitary Waves in Dusty Plasmas with Nonthermal Ions

K. Annou and R. Annou.

*Theoretical physics laboratory, faculty of physics,*
*USTHB, BP 32, El Alia, Bab-Ezzouar, Algiers, Algeria.*

**Abstract.** An investigation of dust-acoustic solitary waves in unmagnetized dusty plasma whose constituents are inertial charged dust grains, Boltzmannian electrons and nonthermal ions has been conducted. The pseudo potential as well as the reductive perturbation methods have been used in high and small amplitude limits. The existence of solitary waves of a positive as well as a negative potential is reported.

## INTRODUCTION

In recent years, there has been a great deal of interest in understanding different types of collective processes in dusty plasmas, which are very common in laboratory and astrophysical environments **[1-4]**. It has been found that the presence of charged dust grains modifies the existing plasma wave spectra, whereas the dust dynamics may even introduce new eigenmodes in the plasma. Indeed, Rao *et al.* **[5]**, were the first to predict theoretically the existence of extremely low-phase velocity dust acoustic waves in unmagnetized dusty plasmas whose constituents are inertial charged dust grains and Boltzmann distributed ions and electrons. These waves have been reported experimentally and their nonlinear features investigated by Barkan *et al.* **[6]**. In this work, we consider the ions non-thermal **[7, 8]** and study their effect on the properties of dust acoustic waves.

## MODEL EQUATIONS

We consider a three-component dusty plasma with extremely massive, micron-sized, negatively charged dust grains, Boltzmannian electrons and nonthermally distributed ions. The quasi-neutrality at equilibrium is written as, $N_{i0} = Z_d N_{d0} + N_{e0}$ where, $N_{i0}$, $N_{e0}$ and $N_{d0}$ are the unperturbed ion, electron and dust densities respectively, and $Z_d$ being the number of elementary charges residing on the dust grain. The electron and nonthermal ions densities are given respectively by,

$$N_e = N_{e0} \exp(\frac{e\phi}{T_e}), \tag{1}$$

$$N_i = N_{i0}\left(1 + \beta\frac{e\phi}{T_i} + \beta\left(\frac{e\phi}{T_i}\right)^2\right)\exp\left\{-\frac{e\phi}{T_i}\right\}, \tag{2}$$

where $\beta = \frac{4\alpha}{1+3\alpha}$, α being a parameter defining the population of nonthermal ions (c.f.Refs.**[7,8]**). Furthermore, the governing equations of the plasma evolution are given by,

$$\frac{\partial N_d}{\partial t} + \frac{\partial}{\partial x}(N_d V_d) = 0, \tag{3}$$

$$\frac{\partial V_d}{\partial t} + V_d\frac{\partial V_d}{\partial x} = -\frac{Z_d e}{m_d}\frac{\partial \phi}{\partial x}, \tag{4}$$

$$\frac{\partial^2 \phi}{\partial x^2} = 4\pi e(N_e - N_i + Z_d N_d), \tag{5}$$

Here, $N_d$, $V_d$ and $m_d$ are the numerical density, velocity and mass of the charged dust grains, respectively.

## FINITE AMPLITUDE LINEAR DUST ACOUSTIC WAVES

We first linearize **Eqs.(1-5)** and assume that the first order quantities of $N_e, N_i, N_d, V_d$ and $\phi$ are proportional to $\exp[i(kx - \omega t)]$ that is $\partial/\partial t = -i\omega$ and $\partial/\partial x = ik$, to get,

$$\tilde{N}_e = N_{e0}\frac{e\phi}{T_e}, \tag{6}$$

$$\tilde{N}_i = N_{i0}(\beta-1)\frac{e\phi}{T_i}, \tag{7}$$

$$-i\omega\tilde{N}_d + ikN_{d0}\tilde{V}_d = 0, \tag{8}$$

$$-i\omega\tilde{V}_d = -ik\phi\frac{Z_d e}{m_d}, \tag{9}$$

$$-k^2\phi = 4\pi e(\tilde{N}_e - \tilde{N}_i + Z_d \tilde{N}_d), \tag{10}$$

The system derived above leads to the following dispersion relation,

$$\frac{\omega}{k} = \frac{v}{\sqrt{1 + \lambda_{Dd}^2 k^2}}, \tag{11}$$

where, $v = \lambda_{Dd} \omega_{pd}$, $\omega_{pd}^2 = \frac{4\pi e^2 Z_d N_{d0}}{m}$, $\lambda_{Dd}^2 = \left(4\pi e^2 \frac{N_0}{T_0}\right)^{-1}$, and $\frac{N_0}{T_0} = \frac{N_{e0}}{T_e} + (1+\beta)\frac{N_{i0}}{T_i}$,

$\lambda_{Dd}$ being the effective Debye length, $\omega_{pd}$ the dust plasma frequency, $N_0$ the effective number density and $T_0$ the effective temperature. For $\lambda_{Dd}^2 k^2 \ll 1$, the dispersion relation reduces to $\omega = kv - \frac{1}{2}\lambda_{Dd}^2 k^3 v$. The argument in the exponent may be written then in the following way,

$$kx - \omega t = (x - vt)k + \frac{1}{2}v\lambda_{Dd}^2 k^3 t, \tag{12}$$

Consequently, it is suggested then to introduce new coordinates $\xi$ and $\tau$ defined by $\xi = \varepsilon^{1/2}(x - v_0 t)$ and $\tau = \varepsilon^{3/2} t$, $v_0$ is the soliton velocity to be determined farther.

## ARBITRARY AMPLITUDE SOLITARY STRUCTURES

In this section we will be looking for arbitrary large amplitude solutions of the nonlinear equations system. The normalised governing equations of the plasma evolution are given by,

$$\frac{\partial n_d}{\partial t} + \frac{\partial}{\partial x}(n_d u_d) = 0, \tag{13}$$

$$\frac{\partial u_d}{\partial t} + u_d \frac{\partial u_d}{\partial x} = \frac{\partial \Phi}{\partial x}, \tag{14}$$

$$\frac{\partial^2 \Phi}{\partial x^2} = n_d + \mu_e n_e - \mu_i n_i, \tag{15}$$

where, $n_d$ is the dust particle number density normalized to $N_{d0}$, $u_d$ is the dust fluid velocity normalized to the dust acoustic speed $c_s = (Z_d T_i / m_d)^{1/2}$, $\Phi$ is the electrostatic potential normalized to $T_i/e$, $\mu = \frac{N_{e0}}{N_{i0}}$, $\mu_i = \frac{1}{1-\mu}$ and $\mu_e = \frac{\mu}{1-\mu}$.

We confine ourselves to investigate stationary solutions that depend on space and time in the following way, $\xi = x - Mt$ (where $M$ is the Mach number). In the stationary frame, we obtain from **Eqs.(13,14)** the density as,

$$n_d = \frac{M}{\sqrt{M^2 + 2\Phi}}, \tag{16}$$

where we have imposed appropriate boundary conditions for the localized disturbances, viz., $u_d \to 0$, $n_d \to 1$, $\Phi \to 0$ at $\xi \to \infty$.

Substituting for $n_d$ from **Eq.(7)** into **Eq.(3)** and multiplying both sides of the resulting equation by $d\Phi/d\mathbf{x}$, integrating once, and taking into account the appropriate boundary conditions, i.e., $\Phi \to 0$ and $d\Phi/d\mathbf{x} \to 0$ at $\xi \to \infty$, we obtain the energy integral equation,

$$\frac{1}{2}\left(\frac{d\Phi}{d\mathbf{x}}\right)^2 + V(\Phi) = 0, \tag{17}$$

where the Sagdeev potential for our purpose reads as,

$$V(\Phi) = M^2\left(1 - \left(1 + \frac{2\Phi}{M^2}\right)^{1/2}\right) - \mathbf{m}\left((1 + 3\mathbf{b} + 3\mathbf{b}\Phi + \Phi^2)\exp(-\Phi) - (1 + 3\mathbf{b})\right) + \frac{\mathbf{m}_e}{\mathbf{s}}(1 - \exp(\mathbf{s}\,\Phi)), \tag{18}$$

where, $\mathbf{s} = \dfrac{T_i}{T_e}$.

The solitonic solutions of **Eq.(18)** exist when the usual conditions, namely, $V(\Phi) = dV(\Phi)/d\Phi = 0$ at $\Phi = 0$ and $V(\Phi) < 0$ for $0 < |\Phi| < |\Phi_0|$, where $|\Phi_0|$ is the maximum amplitude of the solitons, are satisfied. The condition on the Mach number can be obtained by expanding the Sagdeev potential $V(\Phi)$, given by **Eq.(18)**, around the origin. Therefore, the critical Mach number, at which the second derivative changes sign (the lower value of $M$), can be found as,

$$M_l = \sqrt{\frac{1 - \mathbf{m}}{\mathbf{m s} + 3(1 - \mathbf{b})}}, \tag{19}$$

At this critical value of $M$, the third derivative of Sagdeev potential is negative, i.e., both compressive and rarefactive solitary waves exist, in the case we have $\mathbf{a} > \mathbf{a}_{cr}$ ($\mathbf{a}_{cr}$ being a critical value to be determined). Indeed, the condition $V'''(0) < 0$, yields,

$$-3\left(\frac{1-\mathbf{m}}{\mathbf{m s}+3(1-\mathbf{b})}\right)^{-4} - \frac{\mathbf{m}}{1-\mathbf{m}}\mathbf{s}^2 + \frac{1}{1-\mathbf{m}}(7 - \mathbf{s b}) < 0, \text{ which gives } \mathbf{a}_{cr} \sim 0.178. \text{ Furthermore, at}$$

$\Phi_m = -M^2/2$ beyond which the dust density is no longer a positive definite quantity, the Sagdeev potential is required at least to vanish, i.e.,

$$V(\Phi_m) = M^2 - \mu\left(\left(1+3\beta+3\beta\left(-\frac{M^2}{2}\right)+\left(\frac{M^4}{4}\right)\right)\exp\left(\frac{M^2}{2}\right)-(1+3\beta)\right)+\frac{\mu_e}{\sigma}\left(1-\exp\left(-\frac{\sigma M^2}{2}\right)\right) \geq 0, \quad (20)$$

Thus, the upper limit of $M$ (i.e., $M_u$) is that maximum value of $M$ which satisfies the condition $V(\Phi_m)=0$.

We plotted in Figure 1 the lower limit of Mach number $M_l$ versus $\mu$. It is shown that $M_l$ decreases with increasing $\mu$ and $\sigma$. Figure 2 shows how the upper limit of $M$ changes with the parameter $\alpha$. It indicates that as we increase $\alpha$, the upper limit of $M$ also increases. We also plotted the Sagdeev potential versus the electrostatic potential $\Phi$. Figures 3-4 show the variation of Sagdeev potential for $\mu=\sigma=0.01$. It is clear from Fig.3, that for $\alpha > 0.155$ and $1.41 \leq M \leq 1.48$ there is a potential well on the negative $\varphi-axis$, resulting in the existence of rarefactive soliton. Furthermore, it is seen from figures 4 and 5 that for $\alpha = 0.20$, $\alpha = 0.25$ and $1.41 < M < 1.62$ compressive and rarefactive solitons coexist. It has to be mentioned, that solitons correspond to very small hump by opposition to cavitons. As a matter of fact, for $M = 1.55$ (Fig.5) the ratio of the depth of the caviton to the soliton amplitude is around ~10. Figures 6 and 7 depict the variation of Sagdeev potential for a fixed values of the Mach number $M$ and for different value of $\alpha$ and $\mu$. It is found that solitons and cavitons coexist when $\alpha \geq 0.178$ and $\mu < 0.06$. It is interesting to point out that for Boltzmannian ions ($\alpha = 0$), soliton solutions exist for $0.95 < M < 1.52$, and are exclusively rarefactive (c.f.Ref [9]).

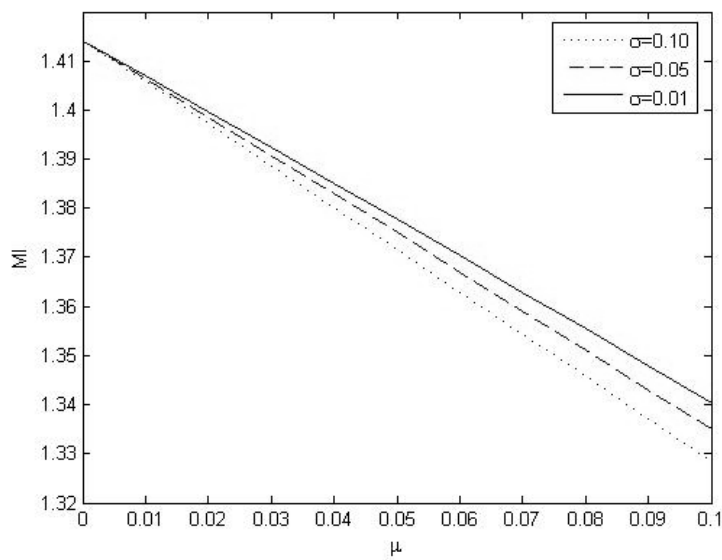

**Fig.1.** $M_l$ vs. $\mu$. It shows how the lower Mach number $M_l$ varies with $\mu$ for $\alpha = 0.20$.

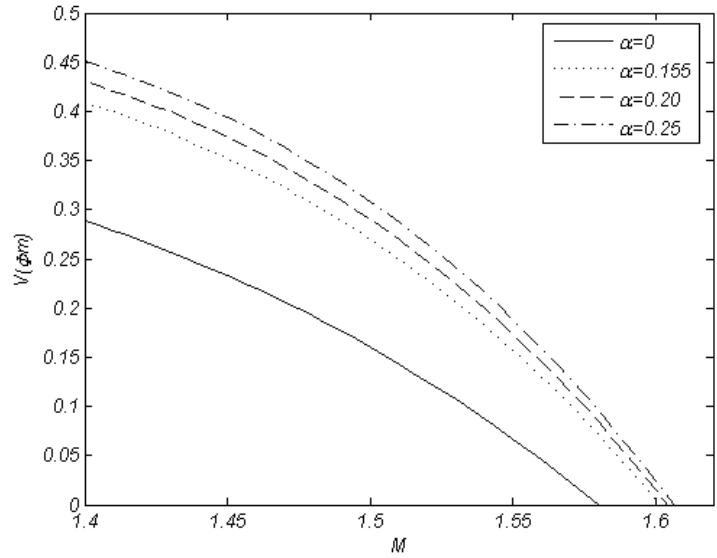

**Fig2.** $V(\Phi_m)$ vs. $M$ for $s = 0.01$ and $m = 0.01$, and different values of $a$.

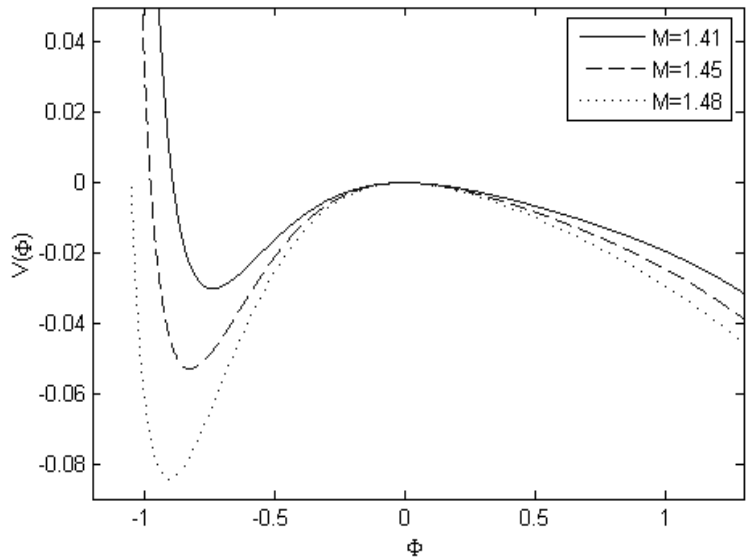

**Fig. 3.** $V(\Phi)$ vs. $\Phi$ for $a = 0.155$. The behaviour of Sagdeev potential $V(\Phi)$ for $a = 0.155$ shows that rarefactive solitons no longer exist when the Mach number exceeds the value 1.48.

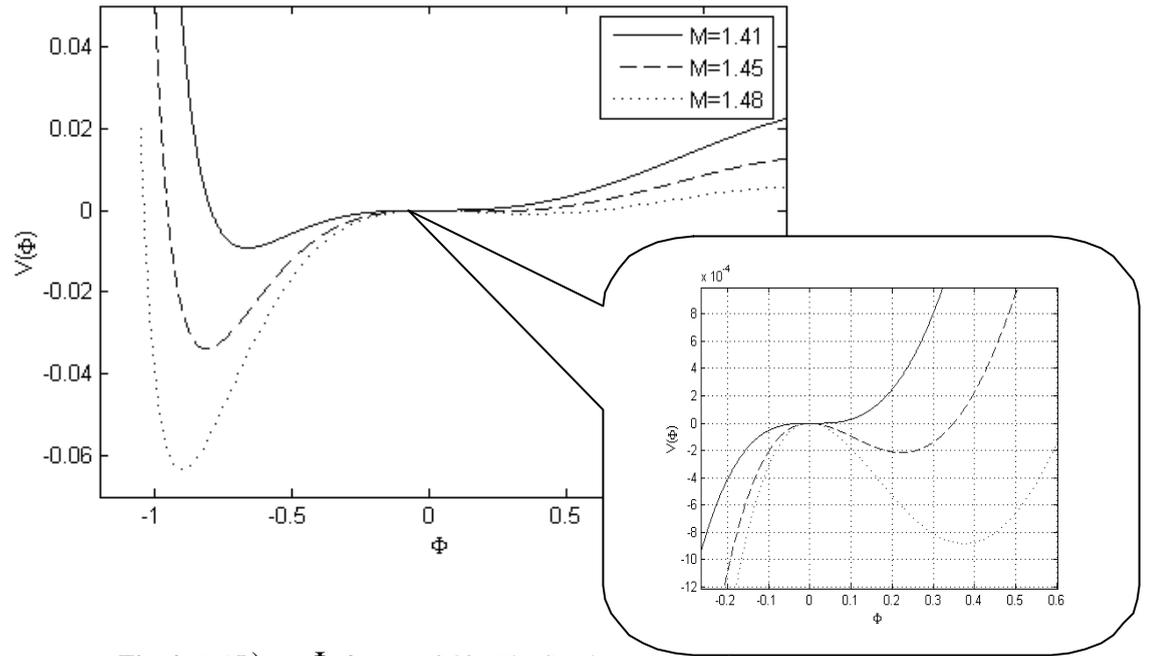

**Fig. 4.** $V(\Phi)$ vs. $\Phi$ for $a = 0.20$. The Sagdeev potential $V(\Phi)$ for $a = 0.20$ shows the coexistence of compressive and rarefactive solitons.

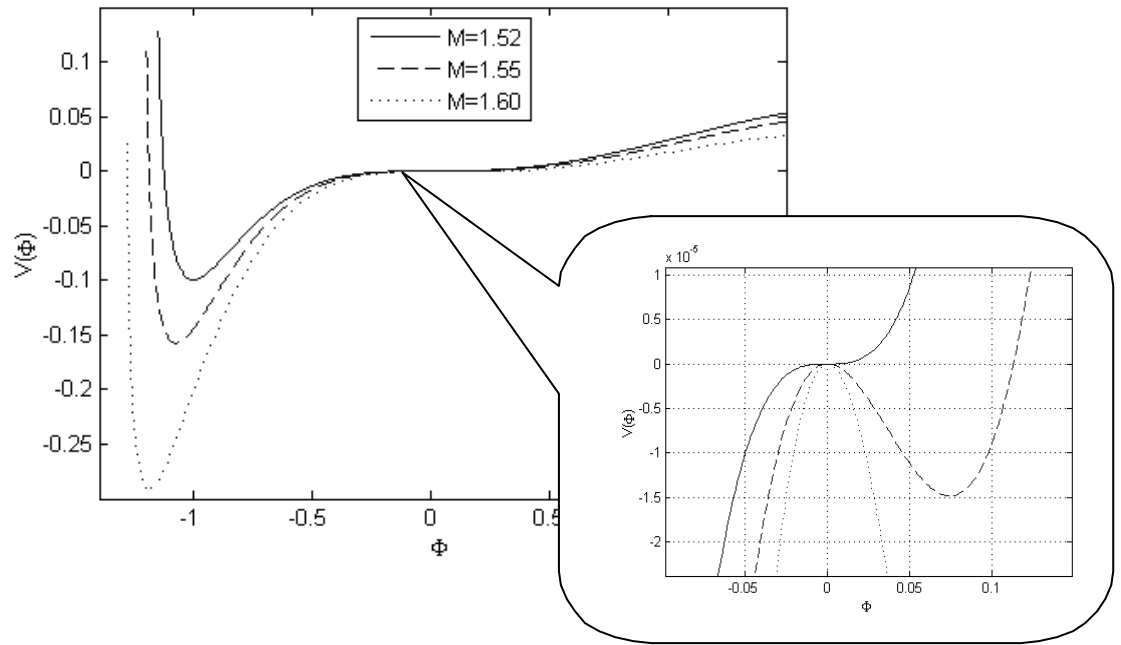

**Fig. 5.** $V(\Phi)$ vs. $\Phi$ for $a = 0.25$. The curve depicts the co-existence of compressive and rarefactive solitons for different values of the Mach number.

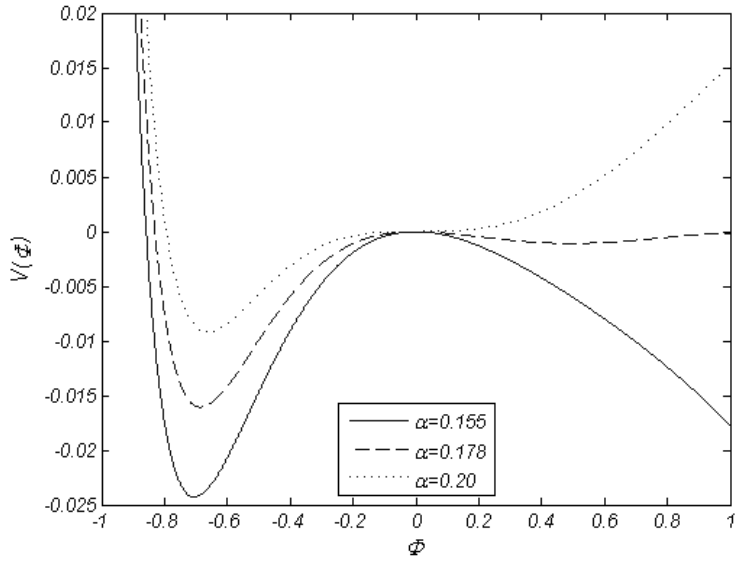

**Fig.6.** $V(\Phi)$ vs. $\Phi$ for $M = 1.41$ and $m = 0.01$. The co-existence of compressive and rarefactive solitons allowed for $a \geq 0.178$.

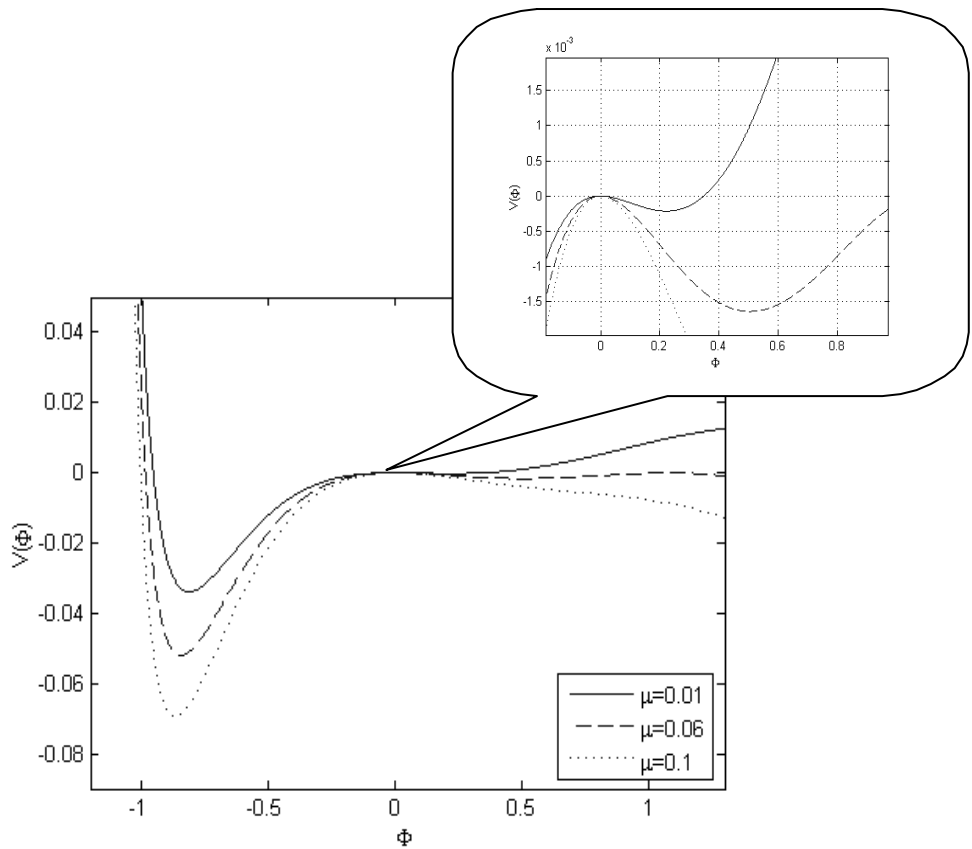

**Fig.7.** $V(\Phi)$ vs. $\Phi$ for $a = 0.20$ and $M = 1.45$. The compressive and rarefactive solitons co-existence is no longer possible when $m$ exceeds the value of $0.06$.

# SMALL AMPLITUDE LIMIT

To study the dynamics of small amplitude dust-acoustic solitary waves, we derive the KdV equation from our basic equations **(13)-(15),** by employing the reductive perturbation method**.** We expand the variables $n_d, m_d$ and $\Phi$ around the unperturbed states in power series of $e$,

$$n_d = 1 + e n_{d1} + e^2 n_{d2} + e^3 n_{d3} + \ldots, \tag{21}$$

$$u_d = e u_{d1} + e^2 u_{d2} + e^3 u_{d3} + \ldots, \tag{22}$$

$$\Phi = e\Phi_1 + e^2 \Phi_2 + e^3 \Phi_3 + \ldots \tag{23}$$

We can rewrite **Eqs.(13-15)** taking into account **Eqs.(21-23)** and the stretched coordinates $x$ and $t$, to get the following system equations,

$$\left(-\sqrt{e}\, v_0 \frac{\partial}{\partial x} + e^{3/2}\frac{\partial}{\partial t}\right) n_d + \sqrt{e}\, n_d \frac{\partial n_d}{\partial x} = 0, \tag{24}$$

$$\left(-\sqrt{e}\, v_0 \frac{\partial}{\partial x} + e^{3/2}\frac{\partial}{\partial t}\right) u_d + \sqrt{e}\, u_d \frac{\partial u_d}{\partial x} = \sqrt{e}\frac{\partial \Phi}{\partial x}, \tag{25}$$

$$e^2 \frac{\partial \Phi}{\partial x} = n_d - m_i[1 + b\Phi + b\Phi^2]\exp(-\Phi) + m_e \exp(s\Phi), \tag{26}$$

To the lowest order in $e$ we have,

$$n_{d1} = -\Phi_1/v_0^2, \tag{27}$$

$$m_{d1} = -\Phi_1/v_0, \tag{28}$$

where, $v_0 = 1/\sqrt{s m_e + m_i(1-b)}$, \tag{29}

To the next order in $e$ we get the following set of equations,

$$\frac{\partial n_{d1}}{\partial t} - v_0 \frac{\partial n_{d2}}{\partial x} + \frac{\partial v_{d2}}{\partial x} + \frac{\partial}{\partial x}(n_{d1} u_{d1}) = 0, \tag{30}$$

$$-v_0 \frac{\partial u_{d2}}{\partial x} + \frac{\partial u_{d1}}{\partial t} + u_{d1}\frac{\partial u_{d1}}{\partial x} - \frac{\partial \Phi_2}{\partial x} = 0, \tag{31}$$

$$\frac{\partial^2 \Phi_1}{\partial x^2} - \frac{1}{v_0^2}\Phi_1 - n_{d1} - g\frac{[\Phi_1]^2}{2} = 0, \tag{32}$$

where, $g = -(m_e s^2 + m_i(1+4b))$,

We can easily eliminate $\partial n_d^{(2)}/\partial x, \partial u_d^{(2)}/\partial x$, and $\partial \Phi_d^{(2)}/\partial x$ from **Eqs.(30-32)**, to obtain the KdV equation,

$$\frac{\partial \Phi_1}{\partial t} + A\Phi_1 \frac{\partial \Phi_1}{\partial x} + B\frac{\partial^3 \Phi_1}{\partial x^3} = 0, \tag{33}$$

where, $A = -\frac{v_0^3}{2}\left[-m_e s^2 - m_i(1+4b) + \frac{3}{v_0^4}\right]$, which may be put in another way, viz.,

$A = -\frac{v_0^3}{2(1-m)^2}\{(1-m)m_is^2 + (1+b)(1-m) + 3(a+m(1+b))^2\}$, and $B = \frac{v_0^3}{2}$.

To discuss the solitary wave solution of **Eq.(33)**, we seek a solution $\Phi_1(\xi)$ which depends on $x$ and $t$ through the variable $z = x - Mt$.

In terms of $z$, **Eq.(33)** can be rewritten as

$$-M\frac{d\Phi_1}{d\zeta} + A\Phi_1 \frac{d\Phi_1}{d\zeta} + B\frac{d^3\Phi_1}{d\zeta^3} = 0, \tag{34}$$

or,

$$B\frac{d}{dz}\left(\frac{d^2\Phi_1}{dz^2}\right) + A\frac{1}{2}\frac{d\Phi_1^2}{dz} - M\frac{d\Phi_1}{dz} = 0, \tag{35}$$

Equation **(35)** may be integrated to yield,

$$\left(A\frac{\Phi_1}{2} - M\right)\Phi_1 + B\frac{d^2\Phi_1}{dz^2} = 0, \tag{36}$$

By multiplying **Eq.(36)** by $\frac{d\Phi}{dz}$, we obtain,

$$B\frac{1}{2}\frac{d}{dz}\left(\frac{d\Phi_1}{dz}\right)^2 + A\frac{1}{6}\frac{d\Phi_1^3}{dz} - \frac{M}{2}\frac{d\Phi_1^2}{dz} = 0, \tag{37}$$

We integrate **Eq.(37)** taking into account the boundary conditions $\Phi_1(z) \to 0$, $\frac{d\Phi_1}{dz} \to 0$, $\frac{d^2\Phi_1}{dz^2} \to 0$, and get,

$$\left(\frac{d\Phi_1}{dz}\right)^2 = \frac{1}{3B}\Phi_1^2(z)[3M - A\Phi_1(z)], \tag{38}$$

On integrating **Eq.(38)**, the solution is found to be of a solitonic form, viz.,

$$\Phi_1(z) = \Phi_{1m} \sec h^2(z/d), \tag{39}$$

where the amplitude $\Phi_{1m}$ and the width $d$ are given by $\Phi_{1m} = 3M/A$ and $d = \sqrt{4B/M}$ with $M > 0$. The propagation of compressive and rarefactive solitons depends on the sign of the coefficient $A$. Equation **(39)** clearly indicates that small amplitude solitary waves with $\Phi > 0$ ($\Phi < 0$) exist if $A > 0$ ($A < 0$). In our case, $A$ can be either positive or negative depending on the parameters $a$ and $m$. Consequently, in our case, viz., for the conditions $M > 1.41$, the plasma can support both types of solitary waves, namely, solitons ($\Phi > 0$) and cavitons ($\Phi < 0$). In fact for $0.155 < a < 0.178$, the existence of rarefactive solitons has been established, whereas beyond the value $a = 0.178$, both types of solitons can be supported by the plasma.

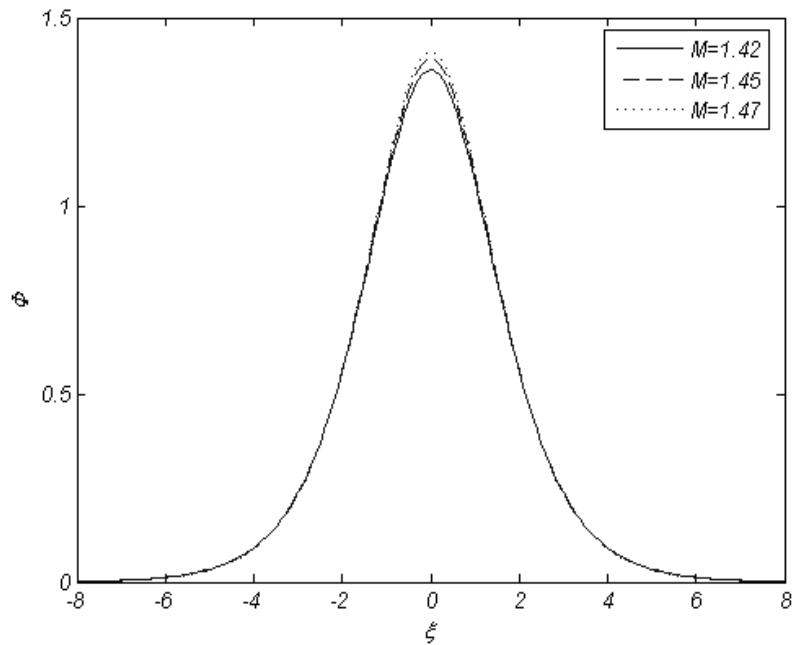

**Fig.8.** Electrostatic potential $\Phi$ versus $z = x - Mt$ for $a = 0.20$, $m = 0.01$ and $s = 0.01$.

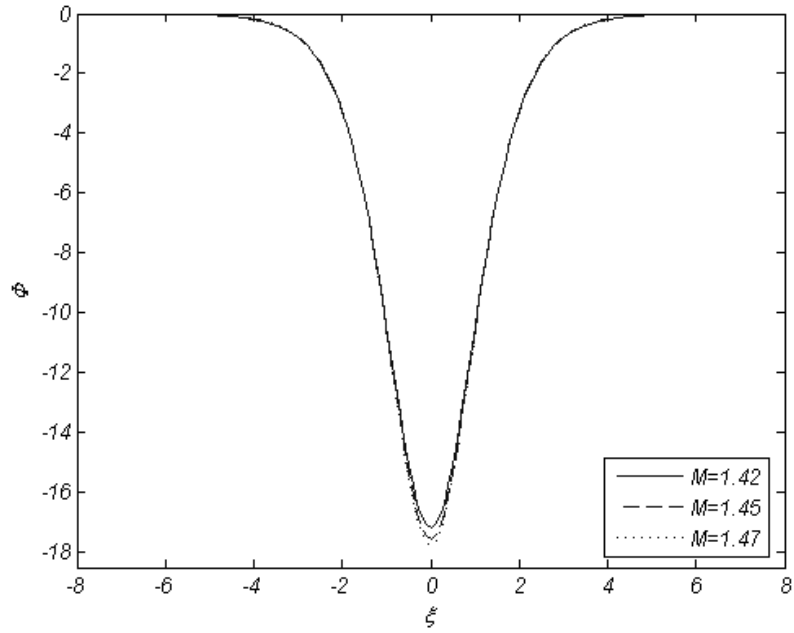

**Fig. 9.** Electrostatic potential $\Phi$ versus $z = x - Mt$ for $a = 0.155$, $s = 0.01$ and $m = 0.01$.

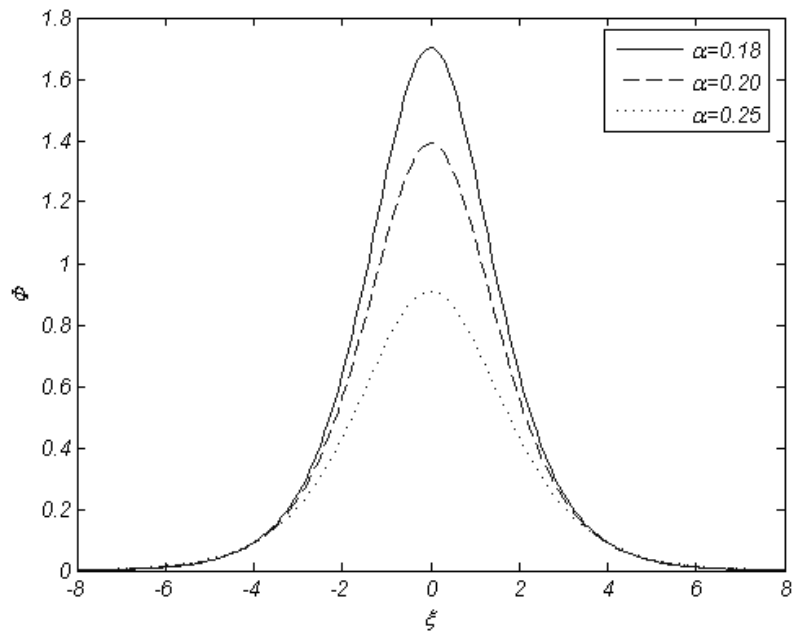

**Fig. 10.** Electrostatic potential $\Phi$ versus $z = x - Mt$ for $M = 1.45$ $m = 0.01$ and $s = 0.01$.

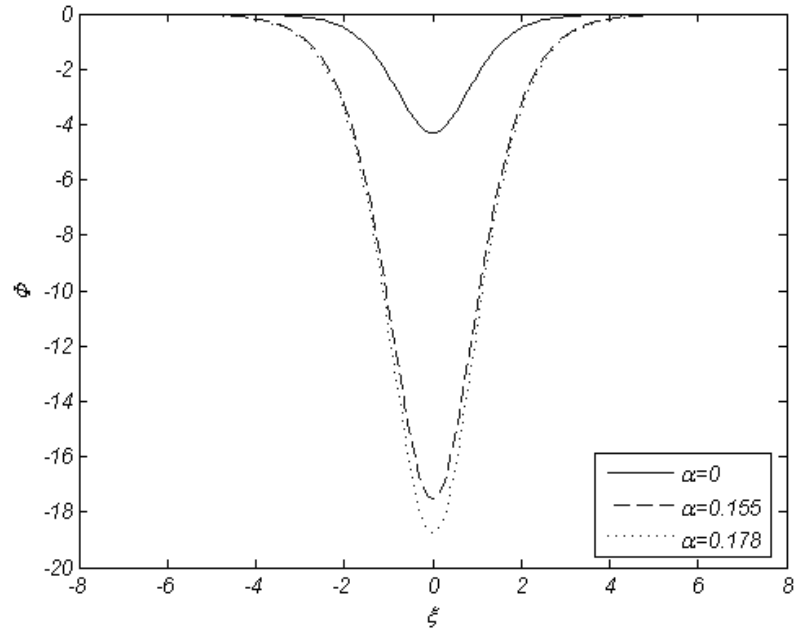

**Fig. 11.** Electrostatic potential $\Phi$ versus $z = x - Mt$ for $M = 1.45$, $\mu = 0.01$ and $s = 0.01$.

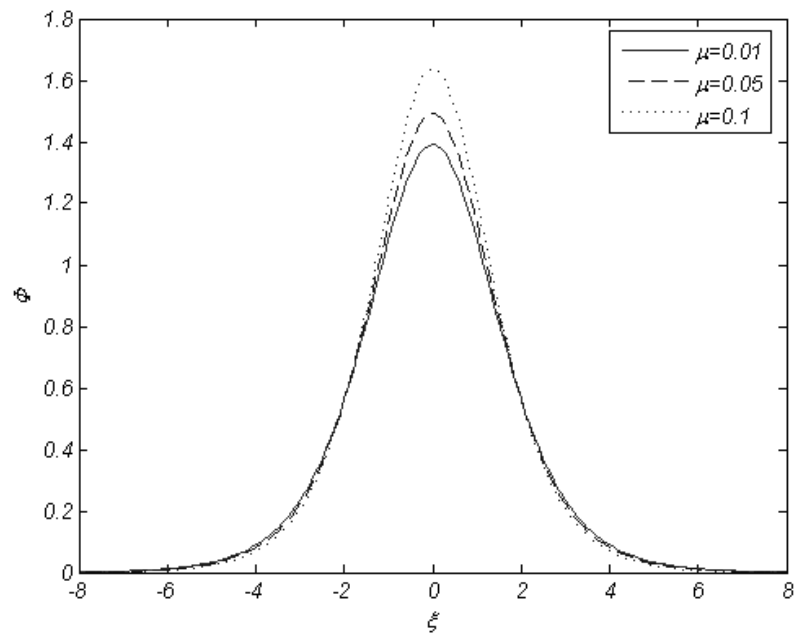

**Fig. 12.** Electrostatic potential $\Phi$ versus $z = x - Mt$ for $M = 1.45$, $a = 0.20$, and $s = 0.01$.

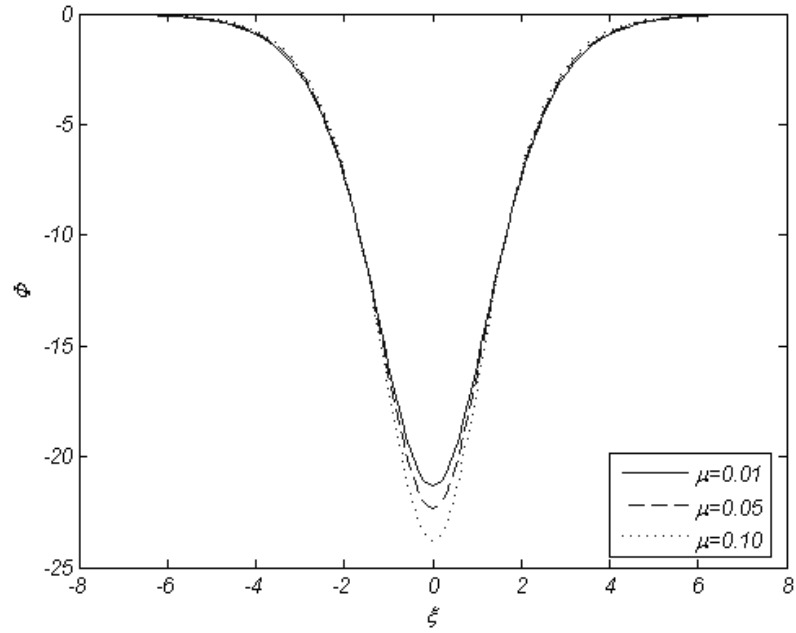

**Fig. 13.** Electrostatic potential $\Phi$ versus $z = x - Mt$ for $M = 1.45$, $a = 0.16$, and $s = 0.01$.

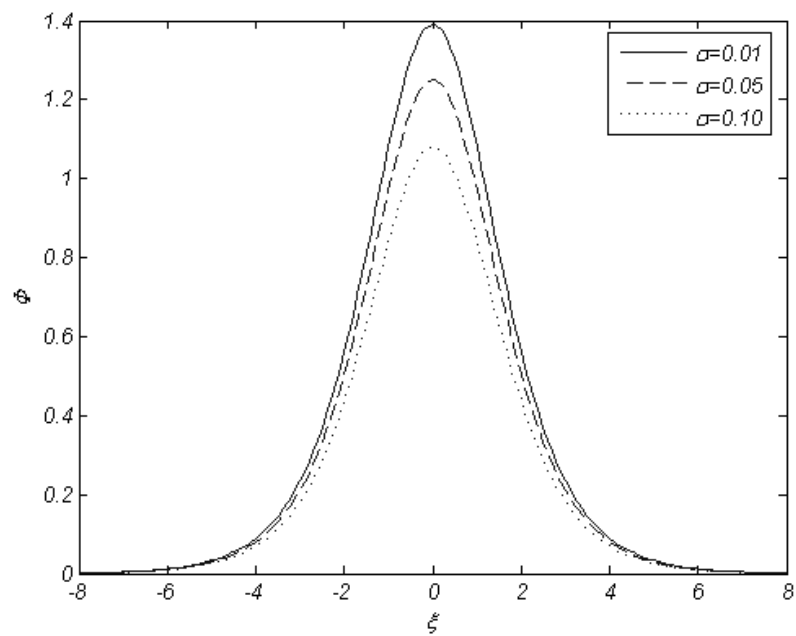

**Fig. 14.** Electrostatic potential $\Phi$ versus $z = x - Mt$ for $M = 1.45$, $a = 0.20$ and $m = 0.01$.

Figures 8-14 show the variation of soliton and caviton profiles for the nonthermally distributed ions for different values of $M$, $a$, $s$ and $m$. It is shown in Figures 8 and 9 that while the soliton amplitude increases with the increase of the Mach number, its width decrease as it was expected. Figures 10 and 11 show that as we increase $a$, the amplitude of the soliton decreases and the depth of the caviton increases, whereas the width of the soliton increases and the caviton width decreases. Moreover, Figures 12-13 show that the width of the soliton decreases with the increase of $m$, while the amplitude increases. Besides, the depth of the caviton increases with increasing the value of $m$, whereas its width decreases. Moreover, it is seen from Figure 14 that the width of soliton decreases when $s$ decreases, while its amplitude increases.

## CONCLUSION

In summary, let us recall that for Boltzmannian ions the soliton solutions exist for $0.95 < M < 1.52$, and that solitons are exclusively rarefactive (cavitons) (c.f. Ref.**[9]**)**.** The condition on the Mach number may be modified if ions are nonthermally distributed. Indeed, the allowed Mach numbers are given by, $1.41 < M < 1.62$ (in case $m = s = 0.01$). Moreover, it is found that inclusion of such a distribution allows the coexistence of solitons and cavitons. Indeed, we have investigated the effect of non-thermal ions on linear and non-linear propagation of dust acoustic waves in unmagnetized dusty plasma consisting of cold dust particles with Boltzmann distributed electrons and non-thermally distributed ions. It is revealed, that as we increase $a$, the upper limit of the Mach number increases. It is found from our study that solitonic solutions exist for $M > 1.41$. Besides, it is interesting to point out that dusty plasmas can support both solitary waves of positive and negative potential corresponding to a dip (caviton) and a hump (soliton) in the dust density, for $a \geq 0.178$ (and $M \geq 1.41$). It is also found that when $0.155 \leq a < 0.178$, we have only rarefactive solitons. Furthermore, a study of solitary waves of small amplitude has been conducted in the framework of the reductive perturbation method. It is revealed that the dynamics of weakly nonlinear and weakly dispersive dust-acoustic waves is governed by a KdV equation. The stationary solution of this equation can be represented in the form of a squared secant hyperbolic. We may mention at the end, that we came to know that Mamun *et al.* also derived the Sagdeev quasi-potential of the plasma under study, which is fortunately the same as the one of **Eq.(18)**; according to their analysis the presence of non thermal ions is favourable to the co-existence of compressive and rarefactive solitons, provided $\alpha > 0.155$ (c.f. Ref. [10]).